\documentclass[aip,apl,preprint]{revtex4-2}
\usepackage{soul}
\usepackage{graphicx}   
\usepackage{dcolumn}    
\usepackage{bm}         
\usepackage[utf8]{inputenc} 
\usepackage[T1]{fontenc}    
\usepackage{mathptmx}       
\usepackage{etoolbox}       
\usepackage{dcolumn}       
\usepackage{supertabular}
\usepackage{siunitx}
\usepackage{xcolor}
\DeclareSIUnit\angstrom{\text {Å}}
\usepackage{array}
\usepackage{xltabular}
\usepackage{makecell}
\usepackage{cleveref}
\usepackage{comment}
\usepackage{xurl}

\begin{document}


\title{Structural and electronic transport investigation of in situ grown InAs-ZnTe core-shell nanowires} 

\author{Danylo Mosiiets}
\altaffiliation{These authors contributed equally to this work.}
\affiliation{Univ. Grenoble Alpes, CNRS, Grenoble INP, Institut Néel, 38000 Grenoble, France}
\author{Chung Wa Shum}
\altaffiliation{These authors contributed equally to this work.}
\affiliation{Department of Physics and Astronomy, University of Pittsburgh, Pittsburgh, Pennsylvania 15260, USA}
\author{Spencer Freeman}
\affiliation{Department of Physics and Astronomy, University of Pittsburgh, Pittsburgh, Pennsylvania 15260, USA}
\author{Martien Den Hertog}
\affiliation{Univ. Grenoble Alpes, CNRS, Grenoble INP, Institut Néel, 38000 Grenoble, France}
\author{Susheng Tan}
\affiliation{Department of Electrical and Computer Engineering and Petersen Institute of NanoScience and Engineering, University of Pittsburgh, PA 15260 USA}
\author{Joël Cibert}
\affiliation{Univ. Grenoble Alpes, CNRS, Grenoble INP, Institut Néel, 38000 Grenoble, France}
\author{Edith Bellet-Amalric}
\affiliation{Univ. Grenoble Alpes, CEA, Grenoble INP, IRIG, PHELIQS, 38000 Grenoble, France}
\author{Sergey Frolov}
\affiliation{Department of Physics and Astronomy, University of Pittsburgh, Pittsburgh, Pennsylvania 15260, USA}
\author{Moïra Hocevar}
\email{moira.hocevar@neel.cnrs.fr}

\affiliation{Univ. Grenoble Alpes, CNRS, Grenoble INP, Institut Néel, 38000 Grenoble, France}

\begin{abstract}
We present a study of ZnTe shells grown in situ around InAs nanowires using molecular beam epitaxy. High-resolution electron microscopy demonstrates epitaxial interfaces, shell thicknesses tunable in the range from 1 to 15 nm, and shells homogeneous along the nanowire axis, thanks to the nearly matching lattice parameters of the shell and the core. The ZnTe shell is insulating, consistent with earlier density functional theory predictions for InAs-ZnTe band edge alignment, and we fabricate electrically contacted field-effect devices by etching through it. Low-temperature transport shows that the field-effect mobility is comparable in core-shell and reference core-only samples, and that the gate hysteresis is reduced in the matched core-shell samples. Ex situ shells show residual oxygen at the interface, despite protective As capping before air exposure, while their mobility and hysteresis remain comparable to those of in situ shells. These results show that ZnTe shells are compatible with field-effect transistor devices, with in situ growth providing the cleanest interfaces, and open the way to hybrid superconductor-semiconductor devices with tunable tunnel barriers, for instance for topological superconductivity. 
\end{abstract}

\pacs{nanowires, field-effect transistor, core-shell, epitaxy, heterostructure, III-V compound semiconductor, II-VI compound semiconductor}

\maketitle 

\section{Introduction}
Combining materials into epitaxial heterostructures had an historical impact on science and technology. Well-known examples include purely III-V semiconductor heterostructures, such as high-mobility GaAs/AlGaAs quantum wells used to discover new transport phenomena~\cite{pfeiffer1989electron}, and InAs/GaAs Stranski-Krastanov quantum dots that are among the brightest single-photon emitters~\cite{StepanovAPL2015}. Purely II-VI interfaces such as CdS/CdTe yielded high-efficiency solar cells~\cite{Wu2004} and room-temperature single-photon emission~\cite{Bounouar2012}; Bose-Einstein polariton condensates were realized in CdTe quantum wells~\cite{Richard2005}. Heterostructures that switch semiconductor families can facilitate deeper understanding of growth mechanisms~\cite{Hocevar2012b} and lead to applications such as optoelectronic hybrid circuits~\cite{Liu2018} and III-V vertical gate-all-around transistors~\cite{Ram2022}. 

However, realizations of such cross-family heterostructures remain scarce, owing to material challenges including interdiffusion and unintentional doping, structural defects from lattice mismatch and crystal phase mismatch, heterovalency, and limited availability of growth equipment. Our study tests the hypothesis that, despite these challenges, such dissimilar material heterostructures could be used to build passivated, high-mobility channels and tailored tunnel barriers for superconducting contacts.
We are interested in InAs-ZnTe III-V/II-VI heterointerfaces for electronic device applications. InAs is a prominent III-V compound semiconductor used in electronics due to high bulk mobility and low contact resistance to metals~\cite{DayerSmall2007}. ZnTe is a II-VI compound semiconductor that has a larger bandgap and, due to its transparency in the visible region and large electro-optical effect, is widely used in terahertz applications~\cite{Wu1996,Bartel2005}. It was proposed that ZnTe can passivate the InAs surface, or act as a tunnel barrier between InAs and a superconductor~\cite{BadawyAdvSci2022,JardineACS2025}.

Here, we report on the molecular beam epitaxy (MBE) growth of ZnTe shells around bottom-up grown InAs nanowires. In situ transfer between the III-V and the II-VI chambers facilitates the formation of abrupt and defect-free epitaxial interfaces. InAs and ZnTe have small lattice mismatch in the zinc-blende phase, and also in the wurtzite phase, which allows the growth of shells with thicknesses up to 15 nm without plastic strain relaxation\cite{Cibert2026}. By locally etching the ZnTe, we fabricate electrical contact to the InAs nanowire core and incorporate the wires into devices. 

We use the field-effect mobility from fits of pinch-off traces with a classical diffusive model. Such fits are unreliable in nanoscale devices where quantum effects are significant, so that mobility only serves as a relative metric within the present study, and not for comparison with other works. In matched pairs of core-only and core-shell samples, the ZnTe shell reduces the gate-voltage hysteresis, while the mobility stays comparable within the standard deviation. For thicker shells, devices are insulating without sufficient etching through the shell, consistent with the Fermi level lying in the ZnTe bandgap, as suggested by band-offset measurements~\cite{Wahi1989} and density functional theory (DFT) calculations~\cite{JardineACS2025}.

Our results indicate that in the future, InAs-ZnTe wires can be combined with a second shell of metallic superconductor, available in the same MBE cluster. Excessive coupling to a metal may substantially increase the free-carrier density in the nanowire, and also suppress the Landé g-factor through wavefunction hybridization with the metal, pushing the system farther from the topological regime~\cite{ReegPRB2018}. In this configuration, ZnTe serves as a tunnel barrier that tailors the superconductor-semiconductor coupling for the realization of Majorana Zero Modes~\cite{Mourik2012,badawy2023}.
 
\section{Results}

For the growth of the InAs core we follow our previously reported protocol~\cite{MosiietsCGD2024}. Vapor-liquid-solid (VLS) growth is performed using Au catalyst particles of different average diameters between 40 and 80 nm, resulting in a distribution of nanowire diameters. Samples are then transferred, either in situ under ultra-high vacuum from the III-V to the II-VI chamber, or ex situ from a stand-alone III-V system to the same II-VI chamber. InAs nnowires grown in the stand-alone system are capped with an amorphous As shell prior to air exposure. ZnTe shells are grown around the InAs nanowire cores at temperatures between 320$^\circ$C and 340$^\circ$C~\cite{Rueda2016}, chosen to limit the vertical ZnTe VLS growth and to allow a significant radial growth. The shell thickness is controlled through the flux from the ZnTe compound effusion cell and the growth time.

The details for the samples studied here are summarized in Table~\ref{shellthickness}, including the amount of material sent to the nanowire sidewalls and deduced shell thicknesses. The  former corresponds to the integrated flux $S$, i.e., the incoming flux multiplied by the growth time, corrected for the angle of the flux projected onto the nanowire sidewall facets. Since the shell thickness $h$ depends on the core radius~\cite{Cibert2026}, we compare the samples using the integrated lateral flux $S$ and the thickness $h$ calculated with the growth model of Ref.~\cite{Cibert2026} for a common total radius $r = r_{core} + h = \SI{35}{\nano\meter}$ (Table~\ref{shellthickness}). This radius corresponds to the nanowires used for the transport measurements discussed below, see supplementary information (SI). Direct TEM measurements of the shell thickness from other individual nanowires are presented in Figs.~\ref{Figure2} and~\ref{Figure3}.

Pairs of core (C) and core-shell (CS) samples C1, CS1 and C6, CS6 are grown on the same day with identical InAs core growth parameters. Shells are grown in situ for CS1 and CS3–CS6, and ex situ for CS2x and CS7x, allowing a comparison between the two approaches (see SI for the complete growth parameters).

\begin{table}[htbp]
\centering
\begin{tabular}{llccc}
\hline
\textbf{Sample} & \textbf{Material} & \textbf{Shell} & \textbf{$S$ (nm)} & \textbf{$h$ (nm)} \\
\hline
CS1 & InAs/ZnTe & in situ  & 3  & 1.7  \\
CS2x & InAs/ZnTe & ex situ  & 21 & 6.9  \\
CS3 & InAs/ZnTe & in situ  & 21 & 6.9  \\
CS4 & InAs/ZnTe & in situ  & 41 & 10.9 \\
CS5 & InAs/ZnTe & in situ  & 0.6  & 0.4  \\
CS6 & InAs/ZnTe & in situ  & 5  & 2.5  \\
CS7x & InAs/ZnTe & ex situ  & 3  & 1.7  \\
C1  & InAs      & -       & -  & - \\
C6  & InAs      & -       & -  & - \\
\hline
\end{tabular}
\caption{List of nanowire samples. $S$ is the integrated lateral flux and $h$ the ZnTe shell thickness, calculated with the growth model of Ref.~\cite{Cibert2026} for a total radius $r = \SI{35}{\nano\meter}$. C1 and C6 are bare InAs cores.}
\label{shellthickness}
\end{table}

\begin{figure}
\includegraphics{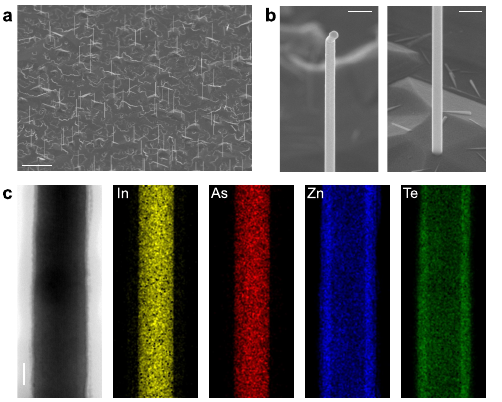}
\caption{\label{Figure1} InAs-ZnTe core-shell nanowires. (a) SEM image of an ensemble of nanowires grown from 40 nm gold colloids (sample CS1). Scale bar 5 $\mu$m. (b) SEM close view of a nanowire (sample CS1). Left: nanowire tip. Right: nanowire bottom. Scale bar 200 nm. (c) BF-STEM image and EDS maps of one nanowire (sample CS2x, ex situ shell) : In in yellow, As in red, Zn in blue, Te in green. Scale bar 20 nm.}
\end{figure}

\begin{figure}
\includegraphics{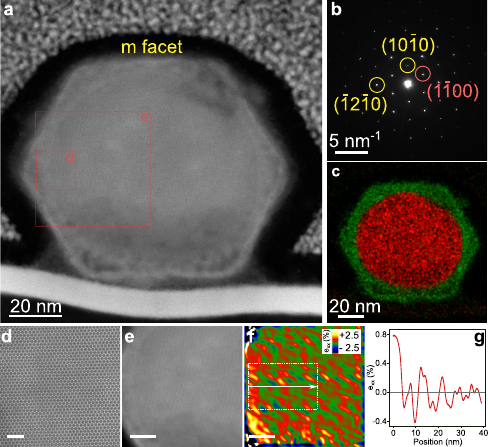}
\caption{\label{Figure2} Epitaxial InAs-ZnTe core-shell nanowires. (a) HAADF-HRSTEM cross section of a nanowire from CS3. (b) SAED pattern of the same cross section wi th indexed diffraction peaks. The indexed (10$\bar{1}$0) and (1$\bar{1}$00) reflections correspond to the ZnTe m-facets in (a), while the ($\bar{1}$2$\bar{1}$0) reflection is aligned with the edge between two ZnTe facets. (c) EDS maps of As (red) and Te (green) from the same cross section. (d) and (e) Higher-magnification images of the regions marked by the red squares in (a) showing the epitaxial interface. Scale bars 2 nm and 10 nm, respectively. (f) Map of the relative variation of the interplanar distance $e_{xx}$ obtained by applying GPA to (e) using the (1$\bar{1}$00) reflection (red circle in (b)). Scale bar 10 nm. (g) Profile along [$\bar{1}$2$\bar{1}$0] of $e_{xx}$ along [$1\bar{1}$00]. Each point is integrated vertically within the rectangle in (f).}
\end{figure}

\begin{figure}
\includegraphics{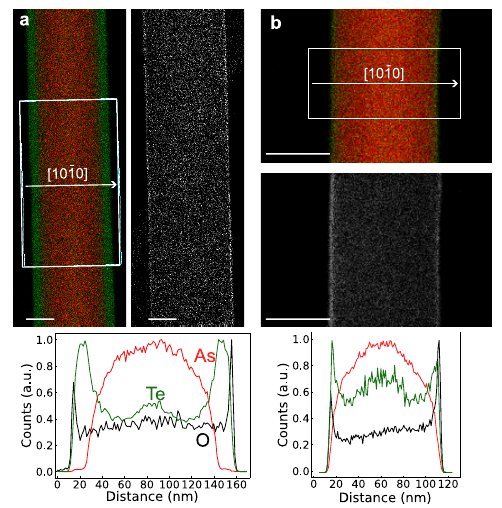}
\caption{\label{Figure3} ZnTe shell thickness control. EDS maps of core-shell nanowires, imaged along the [$\bar{1}$2$\bar{1}$0] zone axis, with corresponding normalized profiles of As (red), Te (green), and O (black). In each panel, the left (top) image corresponds to the As and Te maps, while the right (bottom) image corresponds to solely the O map. (a) Nanowire with a thick shell (CS4), scale bar 50 nm. (b) Nanowire with a thin shell (CS5), scale bar 20 nm.}
\end{figure}

We obtain nanowires with aspect ratios high enough for device applications (Fig.~\ref{Figure1}a). High-resolution transmission electron microscopy (HR-TEM) analysis shows InAs cores in the hexagonal wurtzite crystalline phase (see SI). Low density of unintended defects such as stacking faults and zinc-blende inclusions is found usually in the largest diameter nanowires, yet these defects are also present occasionally in thinner ones. Fig.~\ref{Figure1}b can be used to assess the uniformity of cross-sectional dimensions. The gold catalyst and a short segment of ZnTe nucleated via the VLS mechanism are visible at the tip.

Chemical homogeneity can be evaluated from Bright-Field Scanning Transmission Electron Microscopy (BF-STEM) coupled with energy-dispersive X-ray spectroscopy (EDS) (Fig.~\ref{Figure1}c). The nanowire is from sample CS2x, whose core was grown and capped with As in a stand-alone III-V MBE system, and whose ZnTe shell was grown ex situ. The ZnTe shell is continuous and homogeneous for both in situ and ex situ growth. However, EDS profiles of the nanowire from sample CS2x reveal a trace of oxygen at the InAs/ZnTe interface (see SI), which is absent from in situ shells (Fig.~\ref{Figure3}). 

Ex situ growth also raises morphological issues: after removal of the amorphous As cap, additional ZnTe branches often nucleate on the nanowire sidewalls (see SI). We attribute them either to Au diffusion during decapping, or to In rich surfaces resulting from an aggressive desorption of the cap. Even when ex situ shells are morphologically good, as for CS2x and CS7x, the interface quality is harder to control than with in situ transfer.

For the CS2x nanowire of Fig.~\ref{Figure1}c, we measure a ZnTe shell thickness of 8$\pm$2~nm for an integrated flux $S$ of 21 nm and $r_{core}$ = 14~nm, consistent with the 5.9~nm predicted by the growth model (methodology and
model in the SI). Additionally, under these growth conditions, the shell is uniform along the nanowire axis for all attempted growth times and/or ZnTe fluxes, except when a nanowire core has a high defect density or near the tip \cite{Cibert2026}.

An epitaxial relationship between the shell and the core material translates into fewer misfit dislocations and dangling bonds, which is expected to favor longer electron mean free paths and more uniform tunnel barriers. We investigate the structural quality of the interface between InAs and ZnTe by imaging nanowire cross-sections using high-resolution high angle annular dark field scanning transmission electron microscopy (HAADF-STEM), EDS and geometric phase analysis (GPA) (Fig.~\ref{Figure2}). The InAs cores grown in our chamber generally exhibit a dodecagonal cross section along the entire nanowire length, except at the bottom where the diffusion of In and As adatoms arriving from the substrate stabilizes a hexagonal cross section with a-facets\cite{Mosiiets_thesis}.

The envelope of the nanowire is a hexagon with $\{10\bar{1}0\}$ m-facets (Fig.~\ref{Figure2}a). The orientation is determined from the selective-area electron diffraction (SAED) pattern in Fig.~\ref{Figure2}b. The wurtzite phase of both materials is evidenced in side-view images (see SI). 
A dodecagonal InAs core and a hexagonal ZnTe shell are highlighted in the maps of As and Te EDS signals in Fig. \ref{Figure2}c. All the studied ZnTe shells tend to adopt a hexagonal outer contour at sufficiently large shell thickness, as the faster-growing a-facets are progressively eliminated in favor of the slower growing m-facets~\cite{Cibert2026}.  In this particular nanowire from sample CS3, the shell thickness is \SI{8}{\nano\meter} perpendicular to the $m$-facets, along $[10\bar{1}0]$, and \SI{14}{\nano\meter} at the corners of the hexagon, along $[\bar{1}2\bar{1}0]$, for an InAs core diameter of \SI{74}{\nano\meter}. The shell thickness perpendicular to the m-facet is in agreement with Eq. S1 (see SI). 

In Fig. \ref{Figure2}d, the zoomed-in view shows continuous lattice planes across the interface, confirming the epitaxial relationship. The images also reveal the absence of zinc-blende insertions in this particular cross section, indicated by the absence of dots inside the hexagonal lattice. Note that side-view images are better suited for the observation of dislocations~\cite{Cibert2026}. The low contrast across the interface agrees with the similar average atomic numbers of the elements constituting both materials. 


The crystal lattice is modified across the core-shell InAs-ZnTe nanowire by the change of lattice parameter associated with the change of composition, and by the built-in strain due to the mismatch: the variation of the interplanar distance, for instance $d_{(1\bar100)}$, is the sum of these two contributions. We use GPA to map the relative variation of $d_{(1\bar100)}$ with respect to a reference, which is chosen as the unstrained $d_{(1\bar100)InAs}$ interplanar spacing of the InAs wurtzite phase, though it is possible the InAs core is strained by the ZnTe shell. The relative variation of the interplanar distance is then defined as $e_{xx}$ =$(d-d_{(1\bar100)InAs})/d_{(1\bar100)InAs}$. Within the InAs core, where the composition is constant, $e_{xx}$ isolates the true elastic strain. The relative variation within the ZnTe is the result of both the intrinsic ZnTe-InAs lattice mismatch and the local elastic strain. GPA is performed by applying a mask on the ($1\bar100$) Bragg peak in the Fourier transform (visible in red on the SAED pattern in Fig.\ref{Figure2}b). We choose a mask of size 0.48$\parallel \vec{g} \parallel$ (with $\vec{g}$ the reciprocal vector along the [$1\bar100$] direction).

The GPA map reveals that the interplanar distance is constant in the core within the noise ($e_{xx}$ =-0.03$\pm$0.16\%, mean$\pm$standard deviation) and that $e_{xx}$ increases by about 0.8\% between the InAs core and the ZnTe shell, due to the lattice mismatch and the in-plane strain, including the Poisson effect. In other high-resolution images analyzed by GPA, we observe either no increase of $e_{xx}$ or an increase of 0.5\% to 1\%.  
We compare the maximal value of $e_{xx}$ obtained experimentally with the lattice mismatch calculated with the zinc-blende lattice parameters rather than wurtzite ones, for two reasons: the small differences between the wurtzite and zinc-blende lattice parameters fall below the resolution of our TEM measurements, and the wurtzite values reported in the literature show considerable scatter for both InAs and ZnTe. Using $a_{ZnTe}$ = \SI{6.1037}{\angstrom}~\cite{Madelung2004} and $a_{InAs}$ = \SI{6.0583}{\angstrom}~\cite{Vurgaftman2001}, we obtain a lattice mismatch of 0.75\%, close to the data obtained by GPA. 

We control the ZnTe shell thickness via both growth time and ZnTe flux (Fig. \ref{Figure3}). 
From the normalized EDS profiles of As and Te, we extract a shell thickness of 15$\pm$2~nm for a core diameter of 119$\pm$2~nm in Fig. \ref{Figure3}a for sample CS4. Similarly, we extract a shell thickness of 2$\pm$2~nm for a core diameter of 102$\pm$3~nm from sample CS5 in Fig. \ref{Figure3}b. For nanowires of similar diameter, depositing about 70 times more material on the facets results in shells that are only about 7 times thicker. This factor is much smaller than the flux ratio because the growth rate is not constant: a fast initial growth at small thickness is followed by a slower, nucleation-limited growth~\cite{Cibert2026}, which rate increases with the nanowire radius~\cite{PeetersPRB2025}. Moreover, the growth rate at small $h$ depends on pre-existing nucleation centers, hence on the sample history.
The experimental shell value of the nanowire from CS5 agrees with the thickness of 0.4 nm of the model (Eq. S1 from the SI). For CS4, the model predicts 18.5~nm along the $m$ direction, larger than the measured 15$\pm$2~nm. The discrepancy may arise from two factors: the shell measurement is underestimated because the Te profile does not account for the partial oxidation of the shell, and the nucleation-limited term in the model may overestimate the shell thickness at large radii. Our observations imply that reproducible ultra-thin shells for tunnel-barrier applications will require careful control of the InAs core dimensions. 

No oxygen is detected at the interface of these in situ shells. However, external surface oxidation limits the accuracy of shell thickness calibration for the thinnest shells, on the order of oxide thickness (1-2 nm). Future capping with the second superconductor shell will help alleviate this limitation. The challenge of non-uniform shell thickness around the core, also identified in InAs/CdSe wires~\cite{Kaladzhian2024}, means that the relevant metric for tunnel barriers will be the total tunneling rate through the different facets, rather than a single shell thickness.

\begin{figure}
\includegraphics[width=8.5cm]{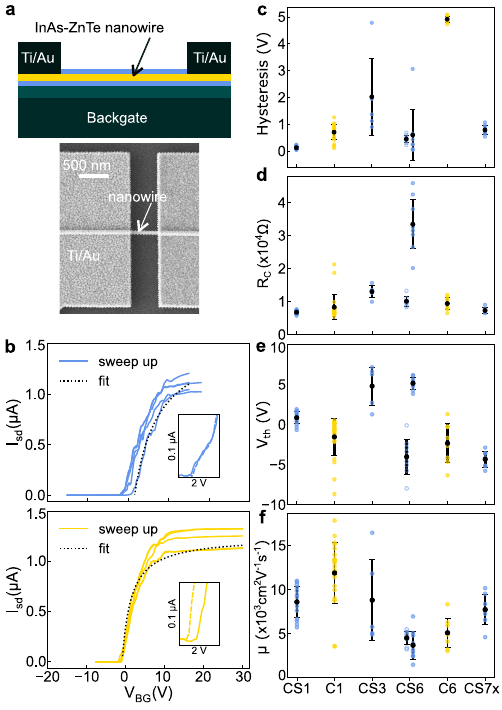}
\caption{\label{Figure4} Transport properties of InAs-ZnTe core-shell nanowires. (a) Schematic side view and SEM top view of a device, with the InAs core in yellow and the ZnTe shell in blue. (b) Pinch-off traces: top, CS1; bottom, C1. The dotted black line in each panel is the fit to Eq. S2 (see SI) of a single trace (one device). In the insets, the solid/dashed lines are sweeps up/down of the gate-voltage zoomed around $V_{th}$ of one device. (c-f) Individual device values, extracted from fits to Eq. S2, for hysteresis, $R_C$, $V_{th}$ and $\mu$. Blue data correspond to the core-shell samples, while yellow data correspond to core-only samples. Open (chip 1) and filled (chip 2) colored circles show devices from different chips of the same sample; black circles with error bars show the sample average and standard deviation. For CS6, devices from the two chips are shown separately, each with its own average. Number of pinched-off devices (see SI): 14 (CS1), 23 (C1), 5 (CS3), 8+12 (chip 1 + chip 2 CS6), 9 (C6), 8 (CS7x, ex situ shell).}
\end{figure}

We now evaluate the impact of the ZnTe shell on the transport properties of the InAs nanowire core. Nanowires are integrated into devices to serve as channels of field-effect transistors. The distance between source and drain contacts is set to \textit{L} = 480 nm (Fig.~\ref{Figure4}a) to reduce the effects of screening from the contacts and minimize short-channel effects. Devices are made using core-shell nanowires from samples CS1, CS3, CS6 and CS7x, as well as core references C1 and C6. For nanowire radii, we aim at an intermediate range around 30-40 nm, since thinner wires tend to not conduct, while thicker wires tend to have structural defects and pinch-off voltages too negative for our measurement setup.

Either the InAs native oxide or the ZnTe shell is removed in the contact regions via dry etching before the deposition of metal contacts (see SI for fabrication methodology). For each sample, pinch-off traces of the source-drain current $I_{sd}$ at finite source-drain bias V$_{sd}$ = 10 mV are obtained for all devices that exhibited pinch-off, using gate-voltage V$_{BG}$ in retrace mode to capture the gate hysteresis (Fig.~\ref{Figure4}b and SI). The pinch-off traces are fit to the formula of Refs.~\cite{GulNanotech2015,BadawyAdvSci2022}:
\begin{equation}
I = \frac{V_{\text{sd}}}{\frac{L^2}{\mu C \left( V_{\text{BG}} - V_{\text{th}} \right)} + R_C},
\label{eq.mob}
\end{equation}
\noindent where $L$ is the channel length, $C$ the capacitance between the nanowire and the back gate, calculated with the model of the core-shell nanowire devices developed in ANSYS Maxwell 3D (see SI), $\mu$ the field-effect mobility, $R_C$ the contact resistance, which limits the current at positive $V_{BG}$, and $V_{th}$ the threshold voltage, below which the current is zero. The gate hysteresis is calculated from the fits of the up- and down-sweep traces, as the difference between the back-gate voltages at which they reach a given current, chosen for each sample (see SI). Fig.~\ref{Figure4}b shows example fits; the extracted parameters are shown in Fig.~\ref{Figure4}c-f; more fits are found in the SI.

By comparing the core-shell samples CS1 and CS6 with their core-only references C1 and C6, we find that the average gate-voltage hysteresis is reduced by 80\% and 90\%, respectively (Fig.~\ref{Figure4}c). This is in contrast with an earlier study where an increase in gate hysteresis was observed when CdTe shells were grown ex situ around InSb nanowires~\cite{BadawyAdvSci2022}. Since gate-voltage hysteresis is typically attributed to the charging and discharging of traps by electrons from the nanowire channel, this reduction indicates that the ZnTe shell lowers the trap density compared with bare InAs. Charge traps are also a source of charge noise, which limits the coherence of semiconductor spin qubits~\cite{DeLeon2021} and nanowire transmons~\cite{Purkayastha2026}. Noise measurements will tell whether ZnTe shells also lead to quieter nanowire devices.

However, devices from the ex situ sample CS7x also show low hysteresis here, leaving open the question of which factors determine the hysteresis. Although the ex situ shells (CS2X, see SI) can be morphologically similar to the in situ ones, their interface contains residual oxygen. Electrically, CS7x devices show hysteresis, mobility and threshold voltage comparable to those of the in situ samples with low contact resistance (CS1, CS6 chip 1). The residual interfacial oxygen therefore does not translate into a measurable difference in these transport metrics.

The average two-terminal contact resistance of the core-shell devices is about \SI{10}{\kohm}, except for CS3 and one of the two CS6 chips, where it is higher (Fig.~\ref{Figure4}d). We extract the threshold voltage at which the devices turn on during a positive gate sweep. On average, the threshold voltage stays within about \SI{\pm5}{\volt}, and in both matched pairs (C1/CS1 and C6/CS6 chip 2), it is more positive for the core-shell samples than for the core-only references (Fig.~\ref{Figure4}e). Devices from one chip (CS6 chip 2, filled circles) show both positive pinch-off voltages and higher contact resistances. This illustrates that not only growth, but also device fabrication affects the extracted parameters. We hypothesize that incomplete dry etching in the contact area explains the higher resistance of this chip, and may also contribute to the higher resistance of CS3, which has a thicker shell. Without the etch step or sufficient etching, the devices show an open circuit, which suggests that the Fermi level is within the bandgap of the ZnTe layer. This is in agreement with DFT calculations on InAs-ZnTe-metal trilayers\cite{JardineACS2025}. 


We report mobility averages in the 4000-9000 \si{\centi\meter\squared\per\volt\per\second} range (Fig. \ref{Figure4}f) for the core-shell nanowire devices. This mobility parameter is extracted from how rapidly the current decreases with gate-voltage. The accuracy is limited by several factors: the recorded gate-voltage range, which may lead to errors in determining R$_C$ and therefore $\mu$ and the difference between the modeled and actual capacitance \cite{BadawyAdvSci2022}. The fitting procedure accounts for some of the variations such as wire diameters and the presence of the shell (see SI). Quantum effects due to confinement and Coulomb blockade can make the gate-voltage dependence steeper; moreover, the quasi-ballistic and few-mode character of low temperature transport is not captured by the fitting formula. 

Again comparing the core-shell samples CS1 and CS6 with their core-only references C1 and C6, we observe that while the presence of the ZnTe shell seems to have a negative impact on the average mobility, the effect is within the standard deviation. For the CS6/C6 pair the values are 4100$\pm$1400 \si{\centi\meter\squared\per\volt\per\second} (average over both chips) and 5400$\pm$1700 \si{\centi\meter\squared\per\volt\per\second}. This is important for applications in topological superconductivity devices, where ballistic transport is a central ingredient. For those applications, the desired mean free paths should ultimately be in the micron range while current mean free paths are likely in the 100 nm range, meaning that both the core and the core-shell need further optimization to reduce scattering.

In conclusion, we achieve epitaxial InAs-ZnTe core-shell nanowires with controlled shell thicknesses down to a few nanometers. The wurtzite ZnTe shell exhibits a slightly larger lattice parameter than InAs. The shells are homogeneous along the nanowire length. However, their thickness varies in the radial direction depending on the crystalline direction, which may lead to different tunneling rates through the different facets. We fabricate field-effect devices by etching through the ZnTe shell. The ZnTe shell reduces the gate hysteresis and the mobility stays comparable within the standard deviation, which makes it a candidate for surface passivation in charge-sensitive devices such as qubits. Compared with ex situ shells grown on As-capped cores, in situ growth avoids interfacial oxygen. Finally, ultra-thin ZnTe shells could serve as tunnel barriers in superconductor-semiconductor devices based on InAs for topological superconductivity.


%
\section{Data Availability}
Raw data of the samples discussed in the present manuscript will be available on Zenodo (SEM images, TEM characterization and pinch-off traces).

\begin{acknowledgments}
Nanowire growth was supported by the IRGA-BICOQ project funded by Université Grenoble Alpes, the ANR PRCI program (QUADIC project ANR-25-CE47-6024), IRP HYNATOQ and the Transatlantic Research Partnership. TEM characterization was supported by the U.S. Department of Energy through grant DE-SC-0019274. Transport measurements were supported by the U.S. Department of Energy, Basic Energy Sciences grant DE-SC-0022073. M. Chernyashevskyy contributed to device fabrication and measurements.
\end{acknowledgments}

\bibliography{biblio_ZnTe_transport}

\end{document}